\documentclass[preprint,authoryear,12pt]{elsarticle}

\usepackage{amssymb, amsmath, subfigure}

\journal{arxiv.org}

\begin{document}

\begin{frontmatter}



\title{Two-timescale evolution on a singular landscape}


\author[CS]{Song Xu}
\ead{song.xu.sjtu@gmail.com}
\author[SCSB,xinyang]{Shuyun Jiao}
\author[Chicago]{Pengyao Jiang}
\author[SCSB]{Ping Ao\corref{cor}}
\ead{aoping@sjtu.edu.cn}

\address[CS]{Department of Computer Science and Engineering, Shanghai Jiao Tong University, 200240, Shanghai, China}
\address[SCSB]{Shanghai Center for Systems Biomedicine, Key Laboratory of Systems Biomedicine of Ministry of Education, Shanghai Jiao Tong University, 200240, Shanghai, China}
\address[xinyang]{Department of Mathematics, Xinyang Normal University, 464000, Xinyang, Henan, China}
\address[Chicago]{Department of Ecology and Evolution, the University of Chicago, 1101 E 57th Street, Chicago, IL 60637, USA}

\cortext[cor]{Corresponding Author}

\begin{abstract}
Under the effect of strong genetic drift, it is highly probable to observe gene fixation or gene loss in a population, shown by infinite peaks on a coherently constructed potential energy landscape. It is then important to ask what such singular peaks imply, with or without the effects of other biological factors. We studied the stochastic escape time from the infinite potential peaks in the Wright-Fisher model, where the typical two-scale diffusion dynamics was observed via computer simulations. We numerically found the average escape time for all the bi-stable cases and analytically approximated the results under weak mutations and selections by calculating the mean first passage time (MFPT) in singular potential peak. Our results showed that Kramers' classical escape formula can be extended to the models with non-Gaussian probability distributions, overcoming constraints in previous methods. The constructed landscape provides a global and coherent description for system's evolutionary dynamics, allowing new biological results to be generated.
\end{abstract}


\end{frontmatter}


\section{Introduction}
\label{Sec:intro}

Interactions between different biological driving forces can generate very complex phenomena in evolution. These forces (genetic drift, mutation, selection, etc.) vary in their forms and intensities of effects, characterized by their different operating timescales \citep{Gillespie1998}. In the study of biological evolution, one of the most important and interesting issues is to describe and separate the multiple timescale dynamics \citep{Gillespie1984} and to analyze the occurring rates of rare events \citep{Kimura1957}. Related problems have been referred in different contexts in literature, in close relation with the concept of adaptive landscape \citep{Wright1932, Arnold2001, Ao2005}. In population genetics, adaptation was found not to be limited by the rate toward local adaptive peaks but by the peak-to-peak transition rate \citep{Wright1932}. Studies were carried on how the success or failure of a mutant gene depends on chance for all levels of selective dominance \citep{Kimura1962}. A knowledge of the frequencies with which populations move toward a local peak and that move between different peaks helps understand the mechanisms of adaptation and divergence \citep{Barton1987}. It was also reviewed how genetic barriers for gene flow are established and related it to biological speciation \citep{Gavrilets2003}. Results on multiple adaptive peaks and the associated multiple evolutionary timescales were found important for studying evolutionary robustness \citep{Ao2009}. The ideas and methodologies are also widely discussed outside biology \citep{Qian2005, Tyll2010}

Typically, the existence of a genetic barrier (or adaptive valley) suggests the separation of different evolutionary timescales. In chemistry, the known Arrhenius formula estimates the separation factor to be an exponential term of the energy barrier height (or valley depth) \citep{Hanggi1990}. It was latter systematically studied in thermally activated systems \citep{Kramers1940}. In population genetics, however, random drift may cause problems for the the classical results. The key point here is that the strength of drift-induced noise varies from state to state and vanishes at the fixation states. This is different from the usual assumption of constant weak noise in physical processes \citep{Gardiner1985}, and the system probability distribution is often far from Gaussian near local equilibria.

To investigate the effects of genetic drift and its interactions with other biological factors, we study the Wright-Fisher diffusion process, a classical model for the study of genetic drift and multi-scale dynamics \citep{Kimura1964, Blythe2007}. It assumes fixed population size and considers the change of portions of different gene copies. In this model, we construct a potential function which is exactly consistent to the potential energy in physical sciences. It can be visualized as a potential landscape, in comparison to the classical fitness landscape \citep{Wright1932}. Under strong genetic drift, a population is very likely to be found near a potential peak at the fixation state. The tendency is shown by singular (infinite) peaks on the potential landscape at such states. The questions here are: What is the biological meaning of these infinite peaks? Do they imply the ultimate fixation of a gene type? If not, what is the life time (average escape time) of such states? In such cases, the classical escape formula gives biologically unexpected estimations for the results. The escape times are often estimated by calculating the numerical solutions of the mean first passage times (MFPT), but the relation between the two concepts was not made clear \citep{Kimura1969, karlin1981second, Lande1985}. In this article, we try to formally find the relation between the escape time and the MFPT in the present model. We look for the analytical estimation of the escape time from the singular potential peaks, and see whether such peaks imply gene fixation.

In more physical contexts, the stochastic escape time in a diffusion process were studied by different methods. One such example is to calculate the stationary probability flux from an attractive basin \citep{Kramers1940}, known as the flux-over-population method. In general diffusion process, this method is shown equivalent to the MFPT calculations \citep{Hanggi1990}. It often assumes Gaussian-like probability distribution near an adaptive peak, which is not the fact in the present model. Another approach is to calculate the eigenvalue of the diffusion equations \citep{Barton1987}. However, the method is not always applicable for bi-stable population models, especially when selection is weak. A third approach is more from the side of population genetics, known as the ``rate of genetic substitution" \citep{Kimura1962, Gillespie1998}. It bypasses the real dynamics in infinite potential peaks, but the results are restricted to the models where fixation is possible and mutation is very weak. In the present article, we study the simulation results of the discrete Wright-Fisher model and approximate the average escape time by calculating the MFPT. Based on previous results \citep{Xu2012}, we study the relation of escape time and MFPT under the present framework. Our results show that Kramers' classical escape formula can be extended to the models with singular potential landscape, overcoming previous constraints and allowing more complex dynamics. Our results provide a complete answer for the bi-stable problems in the present model.

The present article is organized as follows: In Section 2, we introduce the 1-d diffusion process and define a potential landscape. In Section 3, we discuss the uphill and downhill landscape dynamics under strong genetic drift. In Section 4, we first simulate the discrete Wright-Fisher model and analyze the simulated escape rate. We then analytically approximate the average escape time by calculating the MFPT in the infinite potential. After that, we come to two models with selections. In Section 5, we discuss the relation between the MFPT and the escape time. We then compare our results of the escape time and our potential landscape to others' work in literature. To conclude, we discuss the fixation condition in infinite potential and other biological insights based on the present work.

\section{Wright-Fisher model and potential landscape}
\label{Sec:WFmodel}

\subsection{Diffusion process}
\label{Subsec:Diffu}

The 1-d Wright-Fisher model considers the evolution of a diploid population at one locus. The number of individuals in a population is fixed at $N$ and the generations are non-overlapping. Denote the interested pair of alleles as $A_1$ and $A_2$; the total number of the gene copies of $A_1$ and $A_2$ in the population gene pool is $2N$. In the present article, we mainly study the continuous diffusion approximation of the Wright-Fisher model (assume $N$ is big enough for the continuous approximation). Let the frequency of $A_1$ gene be $x$, so the frequency of $A_2$ is $1-x$. Let $\rho(x,t)$ be the probability distribution of $A_1$ at time $t$. The diffusion equation for the continuous Wright-Fisher model is given by \citep{Kimura1964, Ewens2004, Blythe2007}:
\begin{equation}
\partial_t\rho(x,t)=\frac{1}{2}\partial^2_x\Big[V(x)\rho(x,t)\Big]- \partial_x\Big[M(x)\rho(x,t)\Big]~.
\label{Eq:2.1-1}
\end{equation}
$M(x)$ is the average change of the $A_1$ frequency per generation, corresponding to the deterministic factors of the system. $V(x)$ is the variance of the stochastic factors. For example, under mutation and selection:
\begin{equation}
M(x) = -\mu x + \nu (1-x) + \frac{x(1-x)}{2\overline{\omega}}\frac{d\overline{\omega}}{dx}~, \label{Eq:2.1-7}
\end{equation}
where $\mu$ is the mutation rate from $A_{1}$ to $A_{2}$, $\nu$ is that from $A_2$ to $A_1$; $\overline{\omega}$ gives the average fitness of the population in that generation, which depends on $x$. Under random genetic drift:
\begin{equation}
V(x) = x(1-x)/2N ~.
\label{Eq:2.1-8}
\end{equation}
The population size $2N$ (the number of gene copies on the considered locus in the diploid population gene pool) controls the intensity of genetic drift.

In 1-d model, by assuming zero probability current at system equilibrium ($t= +\infty$), the equilibrium probability distribution of Eq.~(\ref{Eq:2.1-1}) can be easily obtained as \citep{Gardiner1985}:
\begin{equation}
\rho(x,+\infty)= \frac{1}{V(x)}\exp\bigg[\int^x \frac{2M(y)}{V(y)}dy\bigg]\bigg/{Z}
= \exp\bigg[\int^x \frac{2M(y)-V'(y)}{V(y)}dy\bigg]\bigg/{Z} ~,
\label{Eq:2.1-5}
\end{equation}
where the normalization constant $Z$ is given by
\begin{equation}
Z = \int^1_0 \exp\bigg[\int^x \frac{2M(y)-V'(y)}{V(y)}dy\bigg] dx~.
\label{Eq:2.1-6}
\end{equation}

\subsection{Potential landscape}
\label{Subsec:PotenLand}

The form of Eq.~(\ref{Eq:2.1-5}) immediately suggests the existence of a potential function $\Phi(x)$ from the Boltzmann-Gibbs distribution in physics \citep{Ao2005}:
\begin{equation}
\rho(x,+\infty)\propto\exp(\Phi(x))~.
\label{Eq:BolzGibbsDistr}
\end{equation}
This definition directly connects the equilibrium probability distribution to a potential energy function, which can be visualized as a potential landscape:
\begin{equation}
\Phi(x)= \int^x \frac{2M(y)-V'(y)}{V(y)}dy \doteq \int^x \frac{f(y)}{D(y)}dy~.
\label{Eq:2.1-4}
\end{equation}
The maxima and minima states on the equilibrium distribution are exactly the peaks and valleys on the potential landscape. Here we have defined a directed force $f(x)$ and an undirected diffusion term $D(x)$, which are closely related to the system's long-term dynamics:
\begin{align}
& f(x)=M(x)-V'(x)/2~, \label{Eq:2.1-3a} \\
& D(x)=V(x)/2~. \label{Eq:2.1-3b}
\end{align}

We can specify Eq.~(\ref{Eq:2.1-4}) in the Wright-Fisher model under the effects of genetic drift, mutation, and selection, by considering Eqs.~(\ref{Eq:2.1-7}) and (\ref{Eq:2.1-8}):
\begin{equation}
\Phi(x)= -\ln x(1-x) + 4N\Big[\nu\ln x+\mu\ln(1-x)\Big]+ 2N\ln\overline{\omega} ~.
\label{Eq:2.1-9}
\end{equation}
With the analytical form of $\Phi(x)$, we may classify the Wright-Fisher diffusion models under different parameters according to their long-term behaviors. Several examples are given in Figure 1.

\section{Bi-stable dynamics under strong genetic drift}
\label{Sec:Bistable}

\subsection{Mutation and genetic drift}
\label{Subsec:MutaDrif}

We apply above methods in the bi-stable cases in the Wright-Fisher model. We start with the simplest mutation-drift case,
\begin{equation}
\Phi(x) = (4N\nu-1)\ln x+(4N\mu-1)\ln(1-x) ~.
\label{Eq:2.1-10}
\end{equation}
To maintain a bi-stable system, we set $4N\nu,4N\mu<1$. There is a unique valley state (saddle point) in $(0,1)$, here we denote as $x=a$:
\begin{equation}
a=(1-4N\nu)/(2-4N\mu-4N\nu)~,
\label{Eq:3-0}
\end{equation}
satisfying $\Phi'(a)=0$ and $\Phi''(a)>0$. Such a valley defines two attractive basins $(0,a)$ and $(a,1)$. A population starting in $(0,a)$ (or $(a,1)$) is expected (averaging the effect of noise) to reach $x=0$ (or $x=1$) peak as a result of directed evolution driven by the directed force $f$. Compared to the usual adaptive force induced by selection, the directed force integrates the effects of other factors (here genetic drift and mutation) besides selection (if there is). Under mutation and drift, there is
\begin{equation}
f(x) = -\mu x+\nu (1-x)-(1-2x)/4N~.
\label{Eq:fMutDrif}
\end{equation}
Obviously there is $f(a)=0$. We will have more detailed discussion on this point in the next section. The potential landscapes under different parameters are shown in Figure 1.

\subsection{Uphill dynamics}
\label{Subsec:Uphill}

The U-shaped bi-stable landscapes in Figure 1 immediately suggest the existence of two-timescale dynamics. The movements of a population, if visualized on the potential landscape, can be classified into two fundamentally different types: uphill and downhill processes, often demonstrated to operate on two different timescales \citep{Zhou2011}. In Figure 2, as an example, a population starting in the $(0,a)$ basin reaches $x=0$ (uphill) in $\mathcal T_1$, which is dominated by the characteristic time of the uphill movements; it then escapes to $x=1$ in $\mathcal T_2$, which is dominated by the downhill movements $0\rightarrow a$. The usual assumption is $\mathcal T_1 \ll \mathcal T_2$.

The uphill valley-to-peak evolution in $\mathcal T_1$ is mainly driven by the directed forces, here our $f(x)$. We refer to the Langevin equation that describes the same evolutionary process with Eq.~(\ref{Eq:2.1-1}), but from the point of view of a single population's stochastic evolution \citep{Ao2007}:
\begin{equation}
\dot{x} = f(x) + \sqrt{D(x)}\zeta(x,t)~,
\label{Eq:SDE}
\end{equation}
Here $\dot{x}=dx/dt$ denotes the change rate of $x$. $\zeta(x,t)$ is the Gaussian white noise. Note that Eq.~(\ref{Eq:SDE}) is related to Eq.~(\ref{Eq:2.1-1}) via a different stochastic integral from those of Ito and Stratonovich \citep{Ao2007}. By averaging the effects of noise over its probability distribution (instead of taking the zero-noise limit $N\rightarrow +\infty$), we obtain the average uphill rate:
\begin{equation}
\dot{x} = f(x)~.
\label{Eq:ODE}
\end{equation}
It is easy to verify that $\Phi$ is non-decreasing along the noise-free evolutionary trajectory of a population:
\begin{equation}
\dot{\Phi} = \Phi'(x)\cdot \dot{x}=f^2(x)/D(x)\geq 0~.
\label{Eq:Lya}
\end{equation}
This manifests Wright's essential idea of non-decreasing scalar function that can be visualized as a potential landscape. For linear $f(x)$, we can always take the approximation form $f\sim -|f|\bar{x}$ (here we replace $x$ with $\bar{x}-a$; note that $\bar x$ gives the distance between $x$ and $a$). The solution of Eq.~(\ref{Eq:ODE}) takes the approximate form
\begin{equation}
\bar{x}(t)= \bar x(0)\cdot\exp(-|f|t)\doteq \bar x(0)\cdot\exp(-t/\mathcal T_1)~,
\end{equation}
where $\bar x(0)$ gives the initial state of the population, and $\mathcal T_1$ is usually called the relaxation time. Under strong genetic drift ($4N\nu,4N\mu\ll 1$), the uphill rate is
\begin{equation}
\dot{x} = f(x) \approx -(1-2x)/4N~.
\label{Eq:2.2-0}
\end{equation}
For $x<1/2$, a population is expected to move toward $x=0$, and vice versa. This is consistent to the biological expectation, that a population is expected to be fixed at either monomorphic state (all the individuals have the same gene type $A_1A_1$, or all are $A_2A_2$) The first time scale for uphill dynamics to reach the local potential peak is
\begin{equation}
\mathcal{T}_1 \sim |f|^{-1} = 2N \cdot \mathcal{O}(1) ~,
\label{Eq:2.2-1}
\end{equation}
the typical operating timescale of genetic drift \citep{Gillespie1998}.

\subsection{Downhill dynamics}
\label{Subsec:downhill}

The downhill dynamics has not an explicit evolutionary trajectory, but is considered as accumulations of rare downhill movements. The process can be characterized by the waiting time $\tau$ for such rare effects to grow big enough, so that a population escapes from the original attractive basin, goes through the potential valley, and stays stable in another attractive basin. In a diffusion model with a finite potential barrier (potential valley), a classical formula estimates the escape time as \citep{Kramers1940}
\begin{equation}
\tau \sim \mathcal T_1 \exp{(\Delta \Phi)}~.
\label{Eq:2.2-4}
\end{equation}
Here $\Delta \Phi$ is the potential barrier height (valley depth). $\exp{(\Delta \Phi)}$ is the Arrhenius term.

For the downhill dynamics in the present case, the classical formula would give an estimation of infinite escape time under strong genetic drift and weak mutation: From Eq.~(\ref{Eq:2.1-10}), there is $\Phi(0)=+\infty$, which leads to $\Delta \Phi = \Phi(0)-\Phi(a)=+\infty$ and thus by Eq.~(\ref{Eq:2.2-4})
\begin{equation}
\tau=+\infty~.
\end{equation}

Under pure genetic drift, this infinite escape time is expected. A population will finally be fixed at either $A_1$ or $A_2$ gene. Without mutation or other input of different genes, the fixation state will not be changed. This is also shown by the stationary distribution as a combination of the Dirac delta functions under pure genetic drift \citep{Mckane2007}:
\begin{equation}
\rho(x,~\infty)=(1-C)\delta(x) + C\delta(1-x)~.
\end{equation}
Here $C = \langle x(t=0) \rangle = \int^1_0 x\rho(x,t=0)dx $ is the initial population state \citep{Mckane2007}. We plot $\Phi$ and $\rho(x,~\infty)$ in Figure 1 (red). In this case, we say that the peak-transition will never happen. The escape time is $\tau=+\infty$.

If, instead, there is additional factors such as (weak) mutation, the infinite escape time may not be a good estimation. Biologically, this mutation-like factor will constantly pull a population away from the monomorphic fixation state. It makes the substitution (peak-shift on the landscape) of a mutant possible, and the probability of such substitutions was often studied \citep{Kimura1962}. Mathematically, Eq.~(\ref{Eq:2.2-4}) causes an unexpected property of the escape time, that $\tau/\mathcal{T}_1$ would change discontinuously with $4N\nu$ (from $+\infty$ to 1 as $4N\nu\rightarrow 1$). We try to obtain better estimation for $\tau$ by simulating the Wright-Fisher model and calculating the MFPT in the next section.

\section{Escape time in infinite potential}
\label{Sec:Escape}

\subsection{Two timescales: distribution's view}
\label{Subsec:DistriView}

To study the stochastic dynamics of a population, one may instead study the evolution of infinite many identical populations. In a diffusion model with two potential peaks, the distribution of the populations is expected to undergo two distinct stages of evolution in two different timescales \citep{Barton1987}. In the first timescale, the probability densities converge to the local potential peaks. Local equilibria are established on different potential peaks in $\mathcal T_1$. In the second timescale, the probability densities transit (escape) between different equilibria. The escape rate of the probability density in an attractive basin is assumed to obey an exponential law \citep{Hanggi1990}:
\begin{equation}
Z_0(t)-Z_0(+\infty) = \Gamma_0 \exp(-\lambda t)~.
\end{equation}
$\lambda$ is the average leaking rate of probability density. Here we define the cumulative probability density in the attractive basin $(0,a)$ as
\begin{align}
Z_0(t) = \int^a_0 \rho(x,t)dx~.
\label{Eq:Z0}
\end{align}
The subscript 0 denotes the variables relating to the dynamical behaviors in $(0,a)$. $\Gamma_0$ is a time-independent function of system parameters. The leaking rate of $Z_0(t)$ changes with time (actually, its change rate depends on the value of $Z_0(t)$ at that time). We may characterize this process by the inverse of the average leaking rate of probability $\mathcal T_2=1/\lambda$, which gives the timescale to establish the global equilibrium.

Note that the leaking rate is the sum of contributions from both attractive basins ($\lambda = \lambda_0 + \lambda_1 = \tau_0^{-1} + \tau_1^{-1}$). Transitions in the two directions will eventually balance each other as $t\rightarrow +\infty$ at the global equilibrium:
\begin{equation}
\tau_0^{-1} \cdot Z_0(t=+\infty) = \tau_1^{-1} \cdot Z_1(t=+\infty)~.
\end{equation}
Here $Z_1(t)=1-Z_0(t)$ is the cumulative probability density in $(a,1)$. The escape time from the $(0,a)$ basin is then given by \citep{Hanggi1990}
\begin{equation}
\tau_0 = \frac{\lambda^{-1}}{1-Z_0(t=+\infty)}~.
\end{equation}

As the present landscape is a biological realization of the physical landscape, we expect the biological population dynamics also obey this two-stage evolution. To verify this, we simulate the change rate of $P(t)$ (a discrete version of $\rho(x,t)$). The results in Figure \ref{Fig:Simulation} shows a clear two-timescale structure of landscape dynamics. With the probability densities initialized in $(0,a)$, the decreasing rate of $Z_0(t)$ is strictly exponential (Figure \ref{Fig:Z0}). Taking the log scale (in the sub-figure), the regressed slope of gives the escape rate toward the global equilibrium ($\lambda=\lambda_0+\lambda_1$). The decay rate shows rigorous exponential distribution, except for the sudden drop at the beginning period of time ($\sim\mathcal T_1\approx 60$ in the example, when the local equilibria has yet to be established). It validates the assumption of the two-scale dynamics, even though the potential peak at $x=0$ is singular and the established local equilibrium is not Gaussian. In the next section, to get an analytical estimation of $\tau$, we study the mean first passage time in infinite potential peak.

\subsection{MFPT in infinite potential}
\label{Subsec:MFPT}

In general diffusion model, one may calculate the mean first passage time from $x_0$ to $x_1$, by referring to the backward equation of Eq.~($\ref{Eq:2.1-1}$). Without loss of generality, we study the stochastic jump out of the attractive basin $(0,a)$. We study a population's MFPT through the valley point $x=a$ to some state $x_1> a$, starting from $x_0\approx 0$ in $(0,a)$. The interested interval is set as $[0,x_1]$, with $x=0$ the reflecting boundary and $x=x_1$ the absorbing boundary \citep{Gardiner1985}:
\begin{equation}
T_\textrm{MFPT}(x_0\rightarrow x_1) =\int_{x_0}^{x_1} \frac{1}{\epsilon D(y)} \exp \big[-\Phi(y)\big]dy\int_{0} ^ y \exp\big[\Phi(z)\big]dz ~.
\label{Eq:3.1-3}
\end{equation}
Here $\Phi$ is the potential landscape in Eq.~(\ref{Eq:2.1-4}). There is no requirement on the configuration of $\Phi$ (finite peak, etc.) when applying Eq.~(\ref{Eq:3.1-3}). However, the MFPT is conceptually different from the escape time \citep{Hanggi1990}. We will show how the MFPT can be used to analytically approximate the escape time in the present model.

In a finite-barrier diffusion process, the escape time was shown to be approximated by the MFPT and takes the form of the Arrhenius exponential factor \citep{Gardiner1985}. Previous approximation methods are mainly established on the following two assumptions: (1) A ``sharp" valley around $x=a$ on the landscape; (2) Gaussian-like probability distribution around $x=0$. However, these two assumptions fail in the present model, as the landscapes have ``fat" valleys and singular peaks under strong genetic drift (Figure 1). The equilibria established near a potential are not Gaussian \citep{Barton1987}. In this section, we take another way to analytically approximate the escape time from the MFPT, in accordance with the landscape configuration in the present model. Under mutation and drift:
\begin{equation}
T_\textrm{MFPT}(x_0\rightarrow x_1) = 4N\int_{x_0}^{x_1} y^{-4N\nu}(1-y)^{-4N\mu}dy \int_0^y z^{4N\nu-1}(1-z)^{4N\mu-1}dz~. \label{Eq:3.1-4}
\end{equation}
The integral term $z^{4N\nu-1}$ near $z=0$ accounts for the infinity of potential peak (and possibly infinite escape time) in the present model. We note that as $4N\nu,4N\mu \rightarrow 0$ the main contribution of the above integral comes from the inner integral in a small interval $[0,y]$ ($y<a$), that is, the incomplete Beta function $B(y;4N\nu,4N\mu)$. Under the same limit, it is numerically shown to be approximated by $y/4N\nu$. Thus the whole integral is approximately of a scale $1/\nu$. More formally, we expand the incomplete Beta function in Eq.~(\ref{Eq:3.1-4}) under $0<1-x_1<1-y<1-z<1$ (expand $(1-z)^{4N\mu-1}$ near $z=0$):
\begin{align}
B(y;4N\nu,4N\mu) & = \int_0^y z^{4N\nu-1}(1-z)^{4N\mu-1}dz~, \notag \\
& = \int_0^y z^{4N\nu-1}\bigg(\sum^{\infty}_{n=0}z^n\prod^n_{k=1}\frac{k-4N\mu}{k}\bigg) dz~, \notag \\
& = \frac{y^{4N\nu}}{4N\nu} + \sum^{\infty}_{n=1}\frac{y^{n+4N\nu}}{n+4N\nu}\prod^n_{k=1} \frac{k-4N\mu}{k}  ~. \label{Eq:3.1-5}
\end{align}
The convergence of the expansion is obvious given $0<y<x_1<1$. Substitute $B(y;4N\nu,4N\mu)$ and expand $(1-y)^{-4N\mu}$ in the outer integral of Eq.~(\ref{Eq:3.1-4}), we obtain
\begin{align}
T_\textrm{MFPT}(x_0\rightarrow x_1) =& \dfrac{x_1-x_0}{\nu} + \dfrac{4N\mu}{\nu} \sum^{\infty}_{n=1}\dfrac{x_1^{n+1}-x_0^{n+1}}{n+1}\prod^n_{k=2}\bigg( \frac{k-1+4N\mu}{k} \bigg) ~ + \notag \\
& 4N (1-4N\mu)\sum^{\infty}_{n=1}\dfrac{x_1^{n+1}-x_0^{n+1}}{(n+1)(n+4N\nu)}\prod^n_{k=2}\bigg( \frac{k-4N\mu}{k} \bigg)  ~.
\label{Eq:3.1-6}
\end{align}
The expansion converges under $\nu>0,~\mu<1/4N$. For the two limiting cases:

(1) $\nu\rightarrow 0$: The expansion of Eq.~(\ref{Eq:3.1-5}) becomes invalid. The leading term of the expansion changes from $y^4N\nu/4N\nu$ to $\ln y$, which becomes sensitive to $x_0$ near 0 then. To ensure the convergence of $T_\textrm{MFPT}(x_0\rightarrow x_1)$ as $x_0\rightarrow 0$, we need $\nu\neq 0$; this is the condition for the escape problem (from $x=0$) to be finite (as we have discussed in Section 3.3). On the other hand, we always have $T_\textrm{MFPT}(0\rightarrow x_1)\rightarrow \infty$ as $\nu\rightarrow 0$.

(2) $\mu\rightarrow 1/4N$: The expansion of $(1-y)^{-4N\mu}$ would not converge for $x_1\rightarrow 1$, as the resulted series would then become a divergent harmonic series. This is also illustrated by the vanishing bi-stability of the system when $4N\mu=1$ (Figure 1, yellow). To ensure the convergence of Eq.~(\ref{Eq:3.1-6}) as $x_1\rightarrow 1$, we need $\mu<1/4N$.

\subsection{Escape time}
\label{Subsec:EscapeTime}

From the results above, we are able to calculate the MFPT between any two points $x_0$ and $x_1$ that satisfy $0\leq x_0 \leq x_1 \leq 1$. The relation between the MFPT and the escape time, however, is not direct. The escape time, defined as the inverse of the average probability rate through the potential valley $x=a$ \citep{Kramers1940}, cannot usually be (intuitively) estimated by $T_\textrm{MFPT}(0\rightarrow a)$ on general potential energy landscape. This is because a population may still have some probability to return back to $x_0$ immediately after its first arrival to $x=a$. In cases with axisymmetric landscape (with axis $x=a$), it is demonstrated that $T_\textrm{MFPT}(0\rightarrow a)$ should be compensated by a factor of 2 when approximating the escape time \citep{Hanggi1990}.

Under $4N\nu,4N\mu\ll 1$, we have by Eq.~(\ref{Eq:3-0}) that $a\approx 1/2$ and the approximately axisymmetric landscape. The escape time $\tau_0$ is (taking $x_0=0$):
\begin{align}
\tau_0&\approx 2\times T_\textrm{MFPT}(0\rightarrow a) \notag \\
&\approx \dfrac{1}{\nu} + \dfrac{4N\mu}{\nu}\sum^{\infty}_{n=1}\dfrac{2^{-n}}{n+1}\prod^n_{k=2}\bigg( \frac{k-1+4N\mu}{k} \bigg)  ~ + \notag \\
&\qquad 4N(1-4N\mu)\sum^{\infty}_{n=1}\dfrac{2^{-n}}{(n+1)(n+4N\nu)}\prod^n_{k=2}\bigg( \frac{k-4N\mu}{k} \bigg)  ~.
\label{Eq:3.2-1a}
\end{align}
Under $4N\nu,4N\mu\ll 1$, the escape time is approximately independent of the initial state $x_0$ in Eq.~(\ref{Eq:3.2-1a}). The escape time is dominated by the leading term of the series $1/\nu$, added by a remaining term of order $4N\mu/\nu$:
\begin{equation}
\tau_0 \approx \nu^{-1}(1+1.23N\mu)~, \label{Eq:3.3-7a}
\end{equation}
The coefficient 1.23 is an approximation of the remaining series in Eq.~(\ref{Eq:3.2-1a}). Under $4N\nu,4N\mu\ll 1$, $\tau_0$ is thus much bigger than the relaxation time ($\sim 2N$) given in Eq.~(\ref{Eq:2.2-1}). This shows the separation of the two timescales and completes our inquiry for the escape time from $(0,a)$.

Another way to look at the MFPT in Eq.~(\ref{Eq:3.1-6}) is to set $x_1=1$ and obtain the substitution time of $A_1$ alleles. It differs from the escape time above by taking into account the dynamical details in the other attractive basin $(a,1)$:
\begin{align}
T_\textrm{MFPT}(0\rightarrow 1)&= \dfrac{1}{\nu} + \dfrac{4N\mu}{\nu}\sum^{\infty}_{n=1}\dfrac{1}{n+1}\prod^n_{k=2}\bigg( \frac{k-1+4N\mu}{k} \bigg)  ~ + \notag \\
&\qquad 4N(1-4N\mu)\sum^{\infty}_{n=1}\dfrac{1}{(n+1)(n+4N\nu)}\prod^n_{k=2}\bigg( \frac{k-4N\mu}{k} \bigg)  ~.
\label{Eq:3.2-1b}
\end{align}
The necessary condition for its convergence ($\nu> 0,~\mu<1/4N$) has been discussed in Section 3.1. In Appendix A we show that the condition is also sufficient. Biologically, we expect that $T_\textrm{MFPT}(0\rightarrow 1)>\tau_0$, as a population would not have to arrive at $x=1$ before it is identified to have escaped from $(0,a)$. This difference is also mathematically demonstrated by Eqs.~(\ref{Eq:3.2-1a})(\ref{Eq:3.2-1b}). Under the limit $4N\nu,~4N\mu\ll 1$, the two equations arrive at the same result $T_\textrm{MFPT}(0\rightarrow 1)\approx \tau_0\approx 1/\nu$. Numerical comparison of $2T_\textrm{MFPT}(0\rightarrow a)$, $T_\textrm{MFPT}(0\rightarrow a)$, our analytical approximation, and escape time simulated from the discrete model are given in Figure \ref{Fig:Escape_u}.

In Figure \ref{Fig:Escape_u}, the escape time is best approximated by $2T_\textrm{MFPT}(0\rightarrow a)$. $T_\textrm{MFPT}(0\rightarrow 1)$ also approximates the results well, but is always bigger. The estimation $T_\textrm{MFPT}(0\rightarrow a)$ is obviously not a good approximation: This is different from the usual model with finite sharp valley, where the escape time is approximated by $T(0\rightarrow x_1)$ with $x_1\approx a$. The simulated escape time is always between $T_\textrm{MFPT}(0\rightarrow a)$ and $T_\textrm{MFPT}(0\rightarrow 1)$. Another observation is that $2T_\textrm{MFPT}(0\rightarrow a)$ is bigger/smaller than the simulated escape time when $4N\nu$ is small/big. The main reason is that the landscape is no longer symmetric when $4N\nu\neq 4N\mu$ (see Section 4.2 for more discussions).

\subsection{Models with weak selection}
\label{Subsec:Sele}

With the effects of selection, the evolutionary rate is usually not linear-dependent on the gene frequency of $A_1$. Its adaptive nature (if works alone) will drive a population monotonously to a fitness peak on the classical fitness landscape. Under mutation and genetic drift, a population is not expected to evolve towards a fitness peak, but towards a potential peak (Eq.~(\ref{Eq:Lya})). We study how the non-linear adaptive selection interacts with mutation and drift on the present potential landscape.

Though $f(x)$ no longer takes the linear form, the first timescale can still be estimated by $\mathcal T_1 \approx 2N \cdot \mathcal O(1)$ under weak selection ($4Ns\ll 1$). This constraint makes sense in that most selections are weak as observed in nature. The general equation for the MFPT when there is mutation, drift and selection is obtained by substituting Eq.(\ref{Eq:2.1-9}) into Eq.(\ref{Eq:3.1-3})
\begin{equation}
T_\textrm{MFPT}(x_0\rightarrow x_1) = 4N\int_{x_0}^{x_1} (1-y)^{-4N\mu}y^{-4N\nu} \big[\overline{\omega}(y)\big]^{-2N} dy \int_0^y (1-z)^{4N\mu-1}z^{4N\nu-1} \big[\overline{\omega}(z)\big]^{2N} dz~.
\label{Eq:3.3-1}
\end{equation}
The inner integral is no longer the standard incomplete Beta function. If we can expand the average fitness term $[\overline{\omega}(y)]^{2N}$ near the singular state $x=0$, an analytical approximation for the MFPT can be obtained by combining the results with Eq.~(\ref{Eq:3.2-1b}).

An example of selection is the symmetric selection \citep{Barton1987}. The fitness setting is $A_1A_1:A_1A_2:A_2A_2=1:1-s:1$. Assuming the Hardy-Weinberg equilibrium, the effect of selective pressure can be described by the allele frequency \citep{Gillespie1998}. The average rate of change in $x$ per generation by selection alone is $M_s(x)=-sx(1-x)(1-2x)$. Here $s$ is the fitness deficit of the heterozygote relative to the homozygote. We have $\overline{w}= 1-2sx+2sx^2$.If further assuming $\mu=\nu$, the potential landscape is exactly axisymmetric to the axis $a=1/2$ (also the potential valley):
\begin{equation}
\Phi(x) = (4N\mu-1)\ln x(1-x)-4Nsx+4Nsx^2 ~,
\label{Eq:3.3-6}
\end{equation}
plotted in Figure 1 (cyan). We study the effects of weak selection on the escape time on the basis of previous discussion. By expanding $e^{-4Nsx+4Nsx^2}$ near $x=0$, the escape rate $\lambda_0=\tau_0^{-1}$ is obtained as
\begin{align}
\lambda_0 &= 1/(2\times T_\textrm{MFPT}(0\rightarrow a)) \notag  \\
&= 4N\int_{0}^{a} e^{4Nsy-4Nsy^2}(1-y)^{-4N\mu}y^{-4N\mu} dy \int_0^y e^{-4Nsz+4Nsz^2}(1-z)^{4N\mu-1}z^{4N\mu-1} dz \notag \\
& \approx \mu/(1+1.23N\mu+0.67Ns)~. \label{Eq:3.3-7b}
\end{align}
Here $a=1/2$ is the saddle (valley) state of the landscape in Eq.~(\ref{Eq:3.3-6}). We fix other parameters, give numerical comparisons among Eq.~(\ref{Eq:3.3-1}) (specified by Eq.~(\ref{Eq:3.3-6}) and take the inverse), Eq.~(\ref{Eq:3.3-7b}) and discrete results in Figure \ref{Fig:Escape_s}. It can be noticed that our approximation also works for $4Ns\approx 1$.

Another typical model considers the asymmetric selection and mutation. Under this condition, the fixation time of a beneficial or a deleterious mutation is often studied. We check the consistency of our results to the previous conclusion, and show that our results can be applied to more general cases with reverse mutation. For example, if we take $s$ as the selective advantage of $A_1$ over $A_2$ ($s\ll 1$), such that $M_s=sx(1-x)$ \citep{Kimura1964}, the average fitness is given by $\overline{w}= 1+2sx~$. We study the average time of substitution of gene $A_1$ by gene $A_2$, which can be approximated by $T_\textrm{MFPT}(0\rightarrow 1)$. Note that to maintain a bi-stable system, we set $1/4N>\mu,\nu$. To take the expansion we further assume $4Ns<1$. Substitute above settings into Eq.~(\ref{Eq:3.3-1}) and obtain:
\begin{align}
T_\textrm{MFPT}(0\rightarrow 1) &\approx 4N \int_{0}^{1} \bigg[ 1+\Big( 4N\mu-4Ns \Big)y \bigg]\bigg[ \dfrac{1}{4N\nu}+\dfrac{1-4N\mu+4Ns}{1+4N\nu}y \bigg]dy ~, \notag \\
& \approx \dfrac{(1+2N\mu-2Ns)}{\nu} ~, \label{Eq:3.3-3}
\end{align}
the substitution time of $A_1$ alleles. From this result, the selective advantage $s$ decreases the substitution time approximately on a linear scale if $4Ns<1$, consistent to the rate of substitution calculated under the same settings without backward mutations ($\mu=0,s\ll 1, 4Ns<1$) \citep{Gillespie1998}:
\begin{equation}
k = \dfrac{1-e^{-2s}}{1-e^{-4Ns}} \times 2N\nu \approx \dfrac{\nu}{1-2Ns}~, \notag
\end{equation}
just the inverse of Eq.(\ref{Eq:3.3-3}) if we take $\mu=0$. If there is reverse mutation ($\mu\neq 0$), the actual fixation would not happen. Our Eq.~(\ref{Eq:3.3-3}) can still be applied in such cases, carrying the meaning of general transition rate in the direction $A_1\rightarrow A_2$.

\section{DISCUSSION}
\label{Sec:Discu}

\subsection{Comparisons with previous work}
\label{Subsec:PrevEscape}

In the present work, we studied the two-timescale landscape dynamics in the Wright-Fisher diffusion model and estimated the escape time from the MFPT. In the 1-d diffusion model, actually, the time-dependent solution $\rho(x,t)$ of the diffusion equation can be analytically obtained \citep{Kimura1964}, and so is the explicit form of escape rate $dZ_0(t)/dt$. However, the exact analytical solution is in very complex form, and the structure of the two-timescale dynamics is very difficult to identify from the complex mathematics. Our landscape-based results, instead, enabled a very clear view of the separation of the two-stage evolution. The typical exponential decay of probability distribution in physics \citep{Hanggi1990} was also observed in this population genetics model. We employed the MFPT method to get an analytical approximation of the escape time $\tau_0$ in the diffusion model, which is also found to be consistent to the eigenvalue estimations in the discrete Wright-Fisher model \citep{Blythe2007}.  We overcame a technical difficulty for applying the classical method into the non-Gaussian equilibria with weak mutations, and further applied it in models with weak selections.

From Eq.~(\ref{Eq:3.3-7a}), the transition time is approximately independent of the population size $2N$. It reminds us of Kimura's known rate formula for the neutral evolution \citep{Kimura1962}, or the rate of neutral substitution: $2N\nu \times 1/2N = \nu$, just the inverse of Eq.~(\ref{Eq:3.3-7a}) if we take $\mu=0$. The inverse of the substitution rate gives the expected time of the appearance of a mutation that is destined to be fixed \citep{Joe2011}. The coincidence of the two results happens under the limit $4N\nu\ll 1$, where the time for the desired mutant to be actually fixed ($\sim 2N$) is negligible. For comparable $\nu$ and $1/2N$, the population size $N$ will have significant effects on the transition rate. Our result Eq.~(\ref{Eq:3.2-1a}) shows how the population size will have effect on the escape time. Moreover, Eq.~(\ref{Eq:3.2-1a}) can further be applied under two-way mutations, which makes the fixation probability of a new mutant (and thus the rate of substitution) incalculable.

Our results show that Kramers' classical escape time results derived from the MFPT or the flux-over-population method can be extended to the non-Gaussian distribution cases. Under $4N\nu,~4N\mu\ll 1$, the result does not show exponential dependency on the valley depth (or barrier height), but rather is controlled by the sharpness of the potential peak (see the sensitivities of Eq.~(\ref{Eq:2.1-10}) and Eq.~(\ref{Eq:3.3-7a}) with respect to $\nu$). The difference is essentially induced by the special types of noise induced by genetic drift. On the other hand, under $4N\nu,~4N\mu\ll 1$, we have $\mathcal T_1\sim 2N$ and $\mathcal T_2\sim \nu^{-1}$; there is still the separation of different timescales, a natural result of our Eq.~(\ref{Eq:3.2-1a}).

The eigenvalue method was used to study the second example in our Section 4.4. The method failed to approximate the transition rate under very weak selection ($s<4\mu$), however, as the approach requires two ``deterministic equilibria" (deterministic equilibrium is defined as a state which would be an equilibrium in the absence of random perturbations; here it can be simply considered as a mutation-selection landscape) \citep{Barton1987}. However, the present Wright-Fisher model has a special form of noise of genetic drift (shown in Eq.~(\ref{Eq:2.1-8})). When the noise is strong, it will result in two ``stochastic equilibria" even when there is only one ``deterministic equilibrium". This is where the eigenvalue method fails.

\subsection{MFPT and escape time}
\label{Subsec:MFPTescape}

In general, an escaped population is assumed to be caught stable in another attractive basin (here $(a,1)$). We do not take into account the cases in which a population comes back immediately after it escapes from the attractive basin $(0,a)$. In models with a narrow potential barrier at $x=a$, the escape time was shown to be approximated by $T_\textrm{MFPT}(0\rightarrow x_1)$ with $x_1>a$ \citep{Gardiner1985}. Here $x_1$ is near $a$. However, this is not the case in the present model, as the potential valley is not narrow. Even after passing through some point $a<x_1<1$, there may still be considerable possibility for it to return to $(0,a)$. In Figure \ref{Fig:MFPT}, the dependence of the MFPT on the end point $x=x_1$ is approximately linear. This conclusion requires that $4N\nu, 4N\mu\ll 1$: From Eq.~(\ref{Eq:3.1-6}), the MFPT is dominated by the first linear term of $x_1-x_0$ under this condition. From this near-linearity, we can see that the MFPT is conceptually very different from the real-time probability flow through each state \citep{Hanggi1990}, which should not be a constant value. For example, the probability flow (in the direction $0\rightarrow 1$) near $x=a$ should be much bigger than that near $x=0$. Comparing with the escape time, the MFPT describes a transient event in the evolution, which has no direct implication for the future dynamics. An example is that, after first reaching $x=a$, there is still $\approx 1/2$ probability to go into $(0,a)$ or $(a,1)$ then (see Appendix B for more detailed discussions).

Figure \ref{Fig:MFPT} also numerically validates the relation $\tau_0=2\times T_\textrm{MFPT}(0\rightarrow a)$. The saddle $x=a$ is a critical state in the population evolution: the most improbable state at long time observation and the unstable fixed point on the potential landscape. In a typical process with Gaussian local equilibria in physics and chemistry, the main escape time is spent in climbing over the saddle point. In the present Wright-Fisher model, the special form of the stochastic term is induced by genetic drift (Eq.~\ref{Eq:2.1-8}), and the landscape configuration is also different from those in the usual cases. The main difficulty of escape lies in overcoming the sharp peak at $x=0$, where most probability densities are concentrated.

Another information that can be read from Figure ~\ref{Fig:MFPT} is that $T_\textrm{MFPT}(0\rightarrow 1)$ is generally bigger than the escape time. This is also easily observed from the comparison between Eq.~(\ref{Eq:3.2-1a}) and Eq.~(\ref{Eq:3.2-1b}). This is because the escape time from $(0,a)$ does not take into account the actual dynamics in the basin $(a,1)$ \citep{Hanggi1990}, and a population does not have to reach $x=1$ before it is considered to have escaped. Eq.~(\ref{Eq:3.2-1a}) is a better approximation.

\subsection{More on our potential landscape}
\label{Subsec:MoreLand}
By theoretical analysis and computer simulations, we show that an evolving population's probability distribution will first follow the uphill direction to the nearest potential peak in $\mathcal T_1$, and then leak downhill through the saddle (valley) state and establish the global equilibrium in $\mathcal T_2$. It verifies that the population dynamics are faithfully described by the present landscape. Its coherency is also demonstrated in the limiting case of pure genetic drift as shown in Section 3.3 and Figure 1 (Green) with $Z=+\infty$. As shown in Section 2.1, $\Phi$ relates to $\rho(x,~t=+\infty)$ through the Boltzmann-Gibbs distribution (if $Z<+\infty$), but is essentially a dynamical description of the system. Its validity does not require a normalizable equilibrium distribution.

The potential landscape Eq.~(\ref{Eq:2.1-4}) can be compared to the classical fitness landscape, which presents only the effects of selection. Other biological factors may generate various evolutionary mechanisms on the fitness landscape without a unified description, along with other controversies \citep{Kaplan2008}. Also, by only taking the measure of fitness, there may be inconsistencies between the dynamics and biology. Under the effects of other forces (e.g. mutation and drift), a population is not always expected to move toward the fitness peak, but toward the potential peak (Section 3.2). One example is the term ``neutral evolution" commonly used in the absence of selection, where different allele-frequency states of a population are not necessarily equally favored by evolution (except the special case $4N\nu=4N\mu=1$), shown in Figure 1. The present potential landscape (as also noticed in literature \citep{Burger2000}) may serve as a substitute for Wright's original landscape that visualizes and quantifies the evolutionary process in a globally coherent way.

An extension to the fitness landscape is the ``deterministic landscape" \citep{Barton1987}, which integrates all deterministic factors of evolution but does not consider genetic drift. The uphill rate on the deterministic landscape is given by $M(x)$ (instead of $f(x)$, compared to our potential landscape). For example, under mutation and random drift, we have $M(x) = -\mu x+\nu (1-x)~$. It has a zero point $a_M=\nu/(\nu+\mu)$, which is also the unique peak state of the deterministic landscape. It fails to describe the Wright-Fisher model when genetic drift is strong ($4N\nu,4N\mu<1$, our potential landscape is bi-peaked); the associated approaches also fail for such cases (see Sections 4.1 and 3.3). Our directed rate $f(x)$ in Eq.~(\ref{Eq:fMutDrif}), instead, gives the correct expectation of directed evolution. The key point here is that genetic drift (noise) also contributes to the directed force in the long-term observations.

Another extension of the classical fitness landscape is the free fitness function \citep{Barton2009} in consideration of the analogy with thermodynamics. It uses the normalization constant $Z$ as a generating function of system's macroscopic information and requires normalizable probability distribution. Moreover, the associated maximum entropy approximation fails under weak mutations. Our present framework does not have certain constraints, and the validity of our landscape construction and the associated approaches is tested in the whole relevant parameter space. It has already been applied in the study of Muller's ratchet \citep{Jiao2012}, a special case where no backward mutations exist.

\subsection{Normalization constants and fixation}
\label{Subsec:Fixation}

By taking $\nu=0$ in Eq.~(\ref{Eq:3.1-6}), we have $\tau_1=+\infty$. No escape is expected to happen once a population ``trapped" into the neighborhood of $x=0$. In Eq.~(\ref{Eq:3.1-3}), the impossibility comes essentially from the infinity of the incomplete Beta function $B(y;4N\nu,4N\mu)$ in Eq.~(\ref{Eq:3.1-5}). More formally, if we define a partial normalization constant for each attractive basin (taking the mutation-drift case as an example) as
\begin{align}
& Z_0 = \int^a_0 x^{4N\nu-1}(1-x)^{4N\mu-1} dx~, \label{Eq:4.3-2a} \\
& Z_1 = \int^1_a x^{4N\nu-1}(1-x)^{4N\mu-1} dx~, \label{Eq:4.3-2b}
\end{align}
Note that $Z_0$ can also be obtained from the definition in Eq.~(\ref{Eq:Z0}) by taking $t=+\infty$: $Z_0 = Z_0(t=+\infty)$. Here $Z_1=1-Z_0$ is similar. The mathematical condition for the biological fixation at $x=0$ (or $x=1$) should be $Z_0=+\infty$ (or $Z_1=+\infty$). When combined with previous discussions, we conclude that $\Phi(0)=+\infty$ (or $\Phi(1)=+\infty$) does not necessarily imply fixation of $A_1$ (or $A_2$) genes. The condition for a population starting from any initial state to be fixed at a monomorphic gene state $x=0$ or $x=1$ is determined by $Z_0=+\infty,~ Z_1<+\infty$ or $Z_0<+\infty,~ Z_1=+\infty$. If $Z_0=Z_1=+\infty$, the fixation will happen at either $x=0$ or $x=1$ on chance. Another observation from the present results is the emerging of absorbing boundaries at the fixation state; the boundary conditions ``artifically" set by \citep{Mckane2007} are more naturally and generally derived then. Note that certain boundary conditions have also been discussed using MFPT in the literature of stochastic theory \citep{karlin1981second}. However, without a proper framework or detailed discussions on the use of MFPT, the results may not be rigorous and may even be erroneous in some cases. A key issue here is the relation between the escape time and MFPT: the MFPT from 0 to 1/2 does not necessary mean the escape from the $(0,a)$ \citep{karlin1981second}, as shown in our Figure \ref{Fig:Escape_u} and \ref{Fig:Escape_s}. Our last comment is that unnormalizable distributions in the diffusion model do not generate real problems for understanding the original discrete model. It instead provides important dynamical and equilibrium information for the understanding of the system. We summarize above conclusions in Table 1.

\begin{table}[h]
\centering
\caption{Summary of the observations in Section 4.3. $Z_0$ and $Z_1$ are the partial normalization constants defined in Eqs.~(\ref{Eq:4.3-2a}) and (\ref{Eq:4.3-2b}). $\tau_0$ and $\tau_1$ are the respective escape times. The ``Absorb-bound." column gives where the absorbing boundary emerges.}
\begin{tabular}{c c c c c c c}
\hline
$Z_0$ & $Z_1$ & $\tau_0$ & $\tau_1$ & Fixation & Absorb-bound.
\\ \hline
$<\infty$ & $<\infty$ & $<\infty$ & $<\infty$ & N/A & Neither
\\ \hline
$=\infty$ & $<\infty$ & $=\infty$ & $<\infty$ & $x=0$ & $x=0$
\\ \hline
$<\infty$ & $=\infty$ & $<\infty$ & $=\infty$ & $x=1$ & $x=1$
\\ \hline
$=\infty$ & $=\infty$ & $=\infty$ & $=\infty$& $x=0$ or $x=1$ & $x=0,1$
\\ \hline
\end{tabular}
\end{table}

\subsection{Comments on the ``stochastic tunneling"}
\label{Subsec:Tunnel}

In study of a three-phase transition problem, a ``stochastic tunneling" effect was termed that allows transition from one state to another, without passing through the middle state \citep{Iwasa2004}. In light of the present framework of potential landscape and discussions of escape time, we commented that ``stochastic tunneling" is a misused term. It is clear that the essential feature of the tunneling effect does not take place here: Tunneling is a quantum mechanical effect. It is tied to the laws of wave mechanics going through under the potential barrier \citep{anslyn2006modern, Ao1994}. A potential barrier (potential valley) is not to be overcome but is tunneled through, which is classically impossible. Furthermore, the tunneling effect is approximately temperature-independent, while the ``stochastic tunneling" disappears when temperature (noise intensity, or here the population size) decreases to zero (population size increases to infinity). The first step of fixation of the deleterious mutation would never happen with zero noise. In fact, this process is just an ``old-type" saddle-passage escape event on an potential landscape. The proper term should be ``thermal activation".

\section*{Acknowledgements}
The critical comments of D. Waxman and T. Kr\"{u}ger on this work are appreciated. We also thank R. S. Yuan, J. H. Shi, Y. B. Wang, and other members in the lab for their constructive comments. We thank X. A. Wang for the technical support. This work was supported in part by the National 973 Project No.~2010CB529200 and by the Natural Science Foundation of China No.~NFSC91029738.

\bibliographystyle{model2-names}
\bibliography{JTB}








\section*{Appendix A. Convergence of Eq.~(\ref{Eq:3.2-1b})}
\setcounter{equation}{0}
\renewcommand{\theequation}{A.\arabic{equation}}

Under $\nu>0$, the convergence of Eq.~(\ref{Eq:3.2-1b}) relies on the convergence of the sum
\begin{equation}
S = \sum^{\infty}_{n=2}\prod^n_{k=2}\bigg( \frac{k-1+4N\mu}{k} \bigg) \frac{1}{n+1} ~. \notag
\end{equation}
We use Raabe's test for series convergence from standard textbooks of real analysis. For $0\leq 4N\mu<1$, we denote
\begin{equation}
c_n = \prod^n_{k=2}\bigg( \frac{k-1+4N\mu}{k} \bigg) \frac{1}{n+1}~. \notag
\end{equation}
Obviously $c_n$ is positive for all $n>0$. First, we have
\begin{equation}
\lim_{n\rightarrow\infty}\frac{c_{n+1}}{c_n}=1~.
\label{Eq:A2-1}
\end{equation}
We then calculate the Raabe terms
\begin{equation}
R_n = n\bigg( \frac{c_{n+1}}{c_n}-1 \bigg) = \big(4N\mu -2\big)\frac{n}{n+2} ~. \notag
\end{equation}
Here $4N\mu-2$ is a constant less than $-1$. By taking the limit $n\rightarrow \infty$,
\begin{equation}
\lim_{n\rightarrow \infty}R_n = 4N\mu -2 < -1
\label{Eq:A2-2}
\end{equation}
The two conclusions in Eqs.(\ref{Eq:A2-1}, \ref{Eq:A2-2}) verify the convergence of the partial sum $S_n$ under $0\leq 4N\mu<1$.

\section*{Appendix B. Interpretation of the factor 2}
\setcounter{equation}{0}
\renewcommand{\theequation}{B.\arabic{equation}}
The choice of factor 2 is because a population will have (on average) 1/2 probability to be caught stable in the $(a,1)$ basin then. This is not always the truth, though, as valley $x_1=a$ is chosen as a perfect absorbing boundary (sink) rather than a smooth distribution of sinks in $(a,1)$ \citep{Hanggi1990}. For the factor of 2 to be exact, we need the valley point to have zero slope ($\Phi(x=a)$) and the landscape to be axisymmetric near the valley. Asymmetry of the landscape far from the valley state will bring higher-order errors to the factor of 2, which is neglected for the present concern and needs further investigations. On the other hand, differences between 2T$_{MFPT}$ and the simulated escape time may also come from the diffusion approximation and expansion of the MFPT.

One detailed interpretation of this factor 2 in a limiting case is given below. Assume that the rates of uphill ($\sim \mathcal T_1$) and downhill ($T_\textrm{MFPT}(0\rightarrow a)$, denoted as $T_a$) movements are well separated (e.g. $4N\nu,4N\mu\ll 1$). Once a population reaches $x=a$ in $T_a$, it has (on average) probability $1/2$ to fall into either attractive basin and immediately ($\sim \mathcal T_1$, assumed to be negligible compared to $T_a$) reaches the corresponding potential peak; once falling back to $x=0$, it will wait another $T_a$ to reach $x=a$ again; it then again has 1/2 chance to reach $x=1$ or return to $x=0$ immediately. Assume this process continues. The expected time to leave $x=0$ can then be obtained as
\begin{align}
\tau&\approx T_a\times\dfrac{1}{2} + 2T_a\times\dfrac{1}{2^2} + \ldots + nT_a\times\dfrac{1}{2^n} + \ldots ~, \notag \\
&= 2T_a ~,
\end{align}
the same result as Eq.~(\ref{Eq:3.2-1a}).

\section*{Appendix C. Discrete Wright-Fisher process}
\setcounter{equation}{0}
\renewcommand{\theequation}{C.\arabic{equation}}

The original Wright-Fisher model is discrete both in time (number of generations) and space (number of copies of $A_1$). It considers the evolution of the probability distribution function $P_t$ (a vector of $2N+1$ elements) with time $t$. The $t$~th generation sampled $2N$ times to give the $t+1$~th generations. The probability that these $2N$ trials of sampling a population with $i$ copies of $A_1$ gene will give $j$ copies of $A_1$ gene is given by the $(i,j)$th element of the transition probability matrix $M$, defined by the binomial distribution:
\begin{equation}
M_{ij}=C_j^{2N}p(i)^j (1-p(i))^{2N-j}~.
\end{equation}
$C_j^{2N}$ is the number of combinations to choose $j$ genes from a gene pool of size $2N$. $p(i)$ is the probability of choosing a $A_1$ gene from the pool. To give the explicit form of $p(i)$, we denote $y=i/2N$. $p(i)$ is determined by the biological factors considered in the model. Here under genetic drift, mutation and selection:
\begin{equation}
p(i)=\frac{(W_{11} y^2+W_{12}y(1-y))(1-\mu)+(W_{12}y(1-y)+W_{22}(1-y)^2)\nu}{W_{11}y^2+2W_{12}y(1-y)+W_{22}(1-y)^2}
\end{equation}
is the probability of sampling an $A_1$ gene in the population gene pool. The system evolves as
\begin{equation}
P_{t+1} = P_t \cdot M
\end{equation}
In the simulation in Section 4.1, the model is under mutation and genetic drift, and the fitnesses are chosen as $W_{11}:W_{12}:W_{22}=1:1:1$. In the first example of Section 4.4, the fitnesses are $W_{11}:W_{12}:W_{22}=1:1-s:1$.

\begin{figure}[h]
\centering
\subfigure[Potential landscapes]{
\includegraphics[scale=0.67]{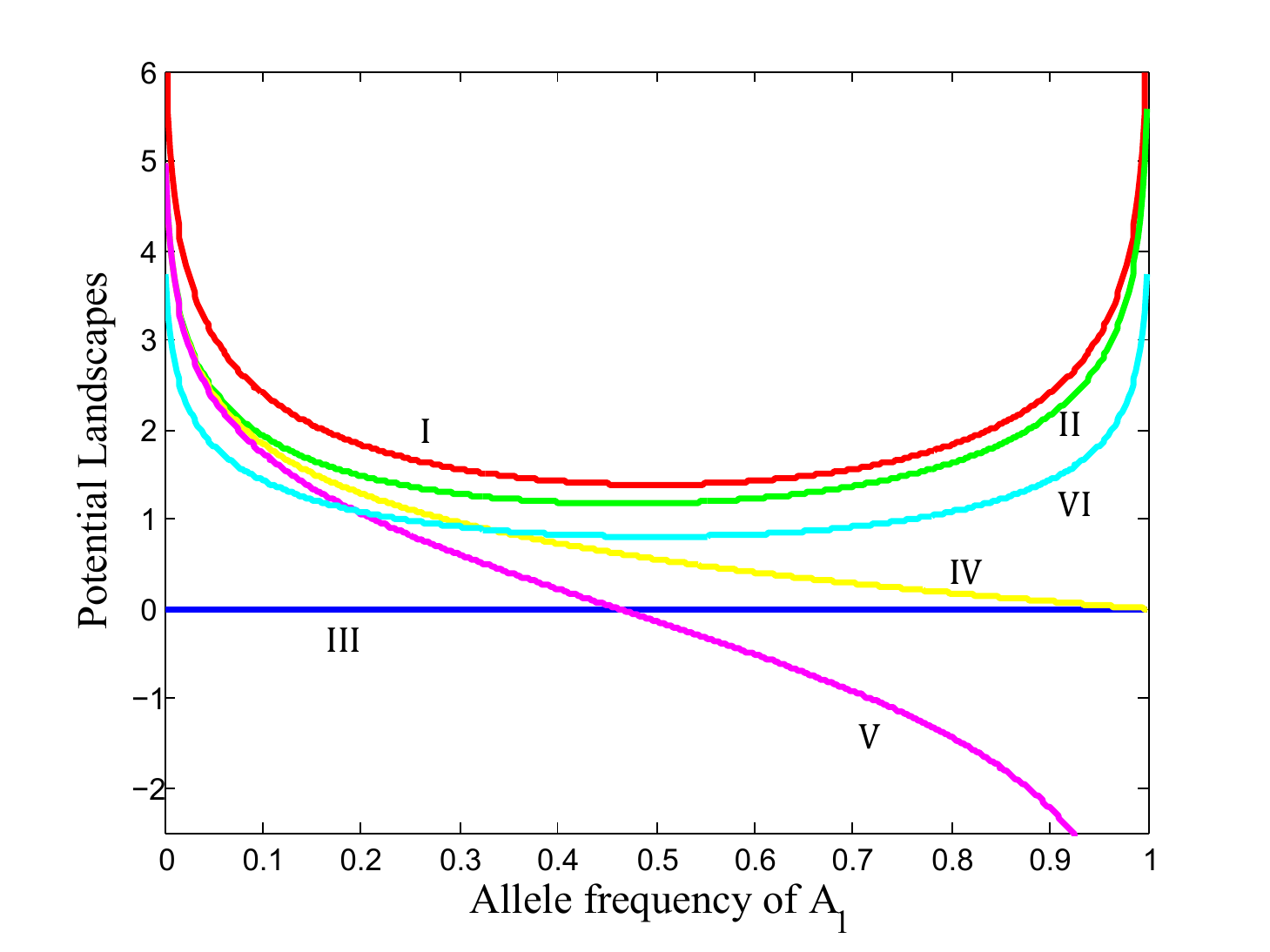}
}
\subfigure[Equilibrium distributions]{
\includegraphics[scale=0.67]{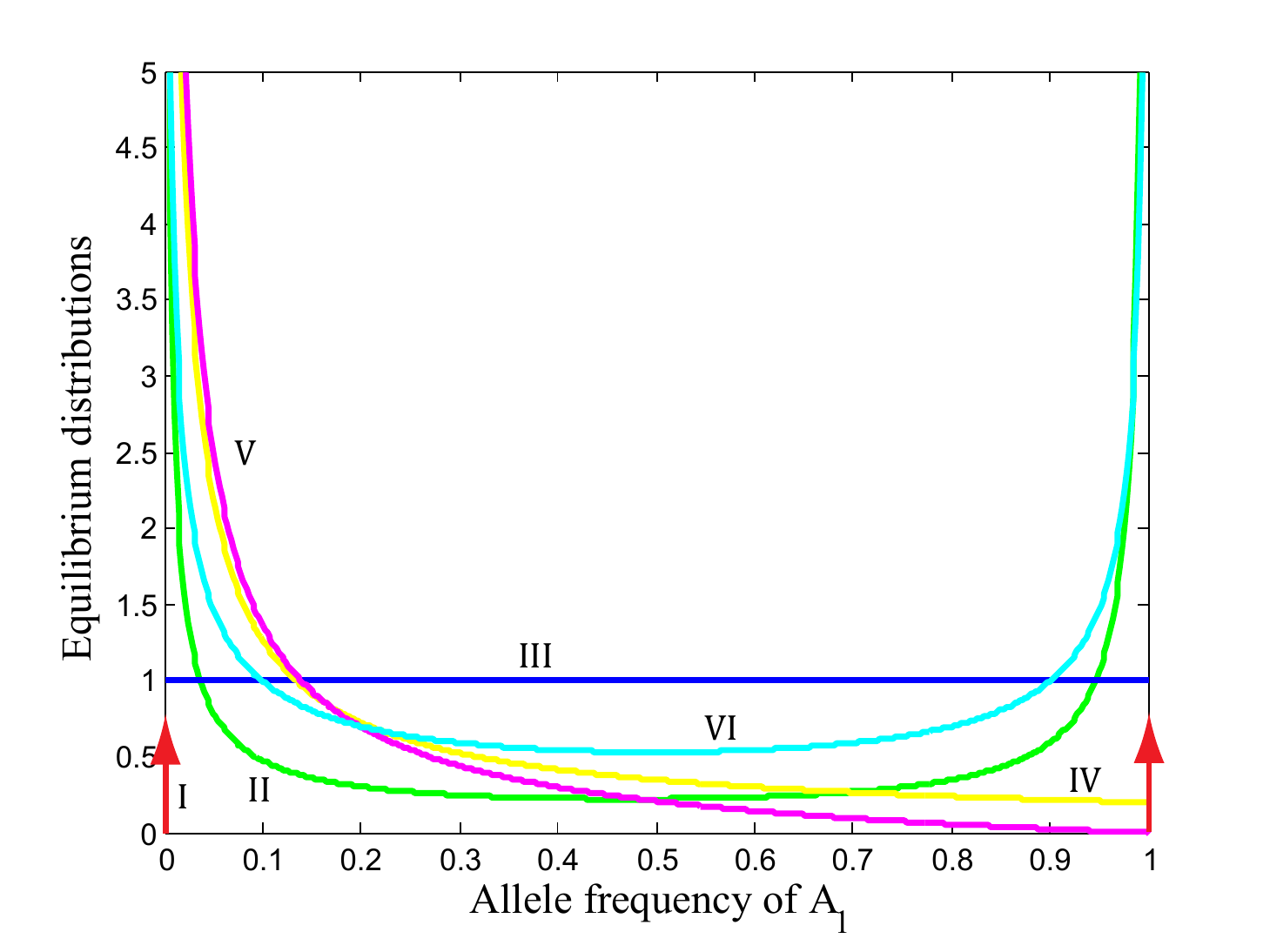}
}
\caption{Potential landscapes and corresponding equilibrium distributions under different parameter settings in the Wright-Fisher model, differentiated by both the colors and Roman indexes. In all cases there is $N=50$. The following five colored landscape contours are generated from Eq.~(\ref{Eq:2.1-10}) under mutation and genetic drift: Red (I): $\mu=\nu=0$. Green (II): $\mu=0.0005,~\nu=0.001$. Blue (III): $\mu=\nu=0.005$. Yellow (IV): $\mu=0.005,~\nu=0.001$. Magenta (V): $\mu=0.01,~\nu=0.001$. The last one is generated from Eq.~(\ref{Eq:3.3-6}), considering mutation, drift, and selection: Cyan (VI): $\mu=\nu=0.002,~s=0.1$. The two red arrows in (b) denote the Dirac delta functions.}
\end{figure}

\begin{figure}[h]
\centering
\includegraphics{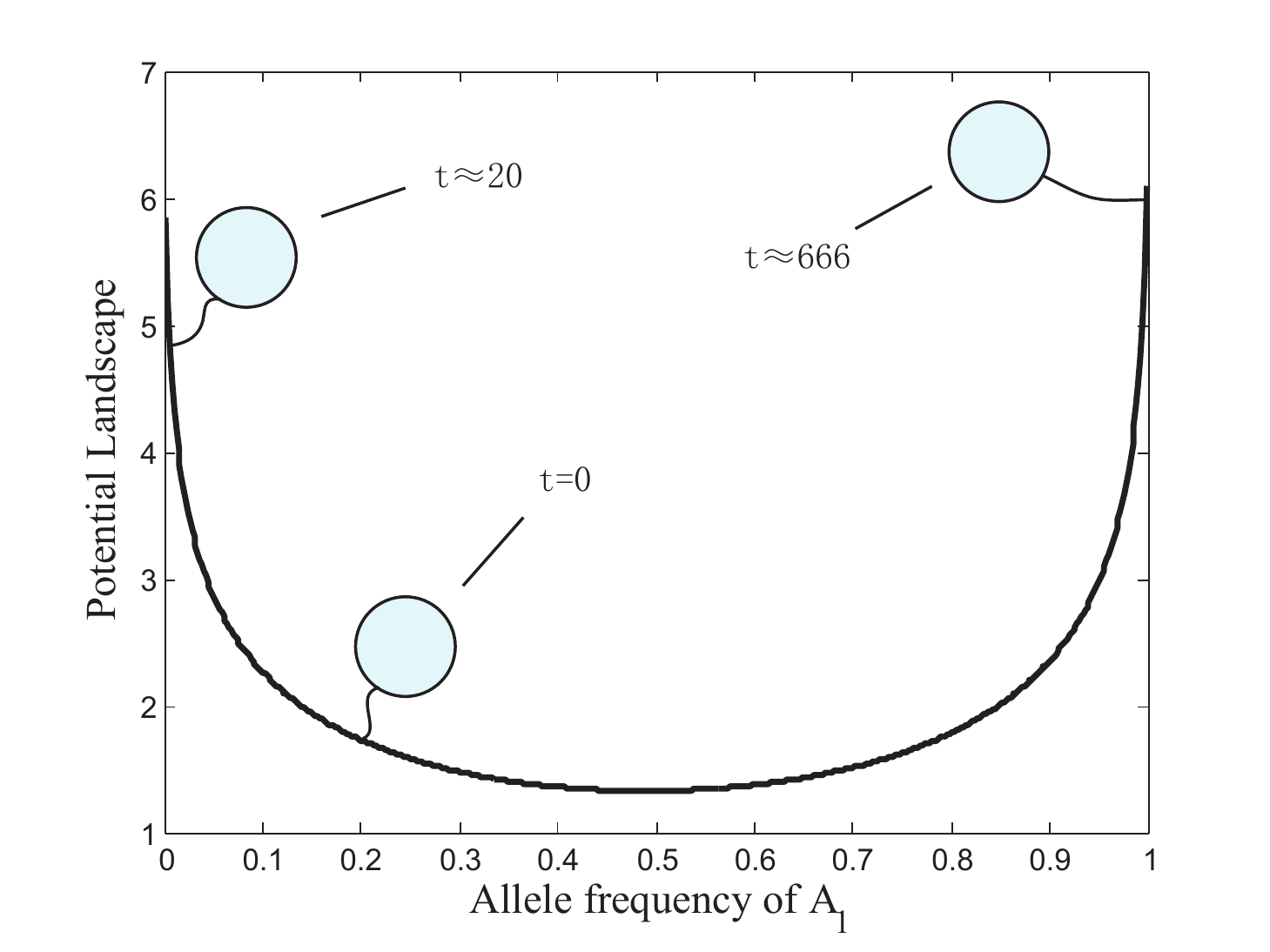}
\caption{Visualization of the two-scale dynamics on a typical bi-stable landscape in the present model. It gives the most probable state of a population (denoted as a balloon, which always searches for a higher ``altitude" to stay) in different timescales visualized on the potential landscape. The parameters are:  $2N=20,~\mu=0.0005,~\nu=0.0015$.}
\label{Fig:Cartoon}
\end{figure}

\begin{figure}[h]
\centering
\includegraphics{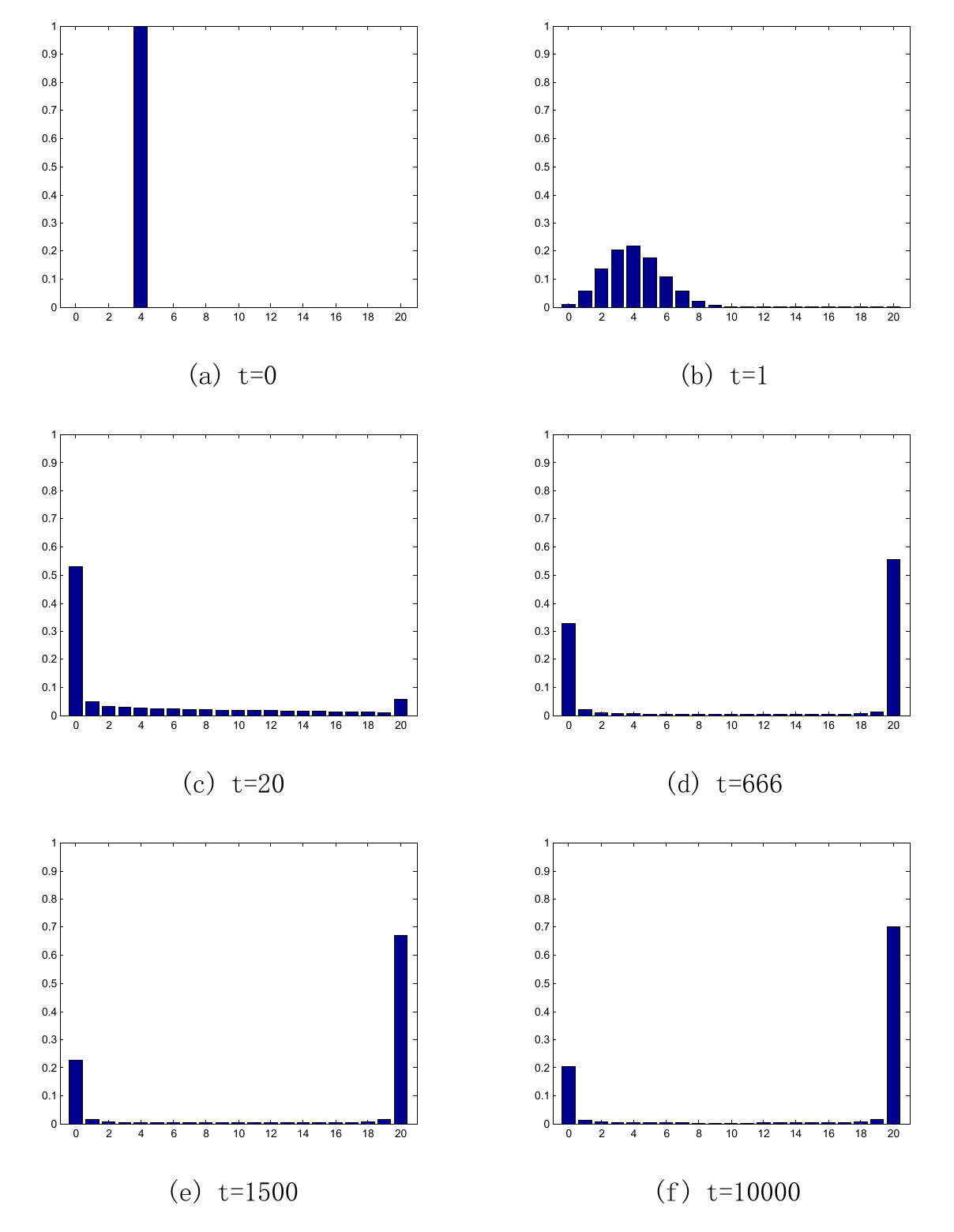}
\caption{Simulations of the discrete Wright-Fisher model under mutation and random drift. x-axis gives the number of $A_1$ alleles and y-axis is the probability distribution. Parameter settings: $2N=20,~\mu=0.0005,~\nu=0.0015$, so that $T_1\approx20,~\tau_0\approx 666$. (a) shows that the initial state is set to $x=0.2$. (e) and (f) show the establishment of the equilibrium distribution after long enough time.}
\label{Fig:Simulation}
\end{figure}

\begin{figure}[h]
\centering
\includegraphics{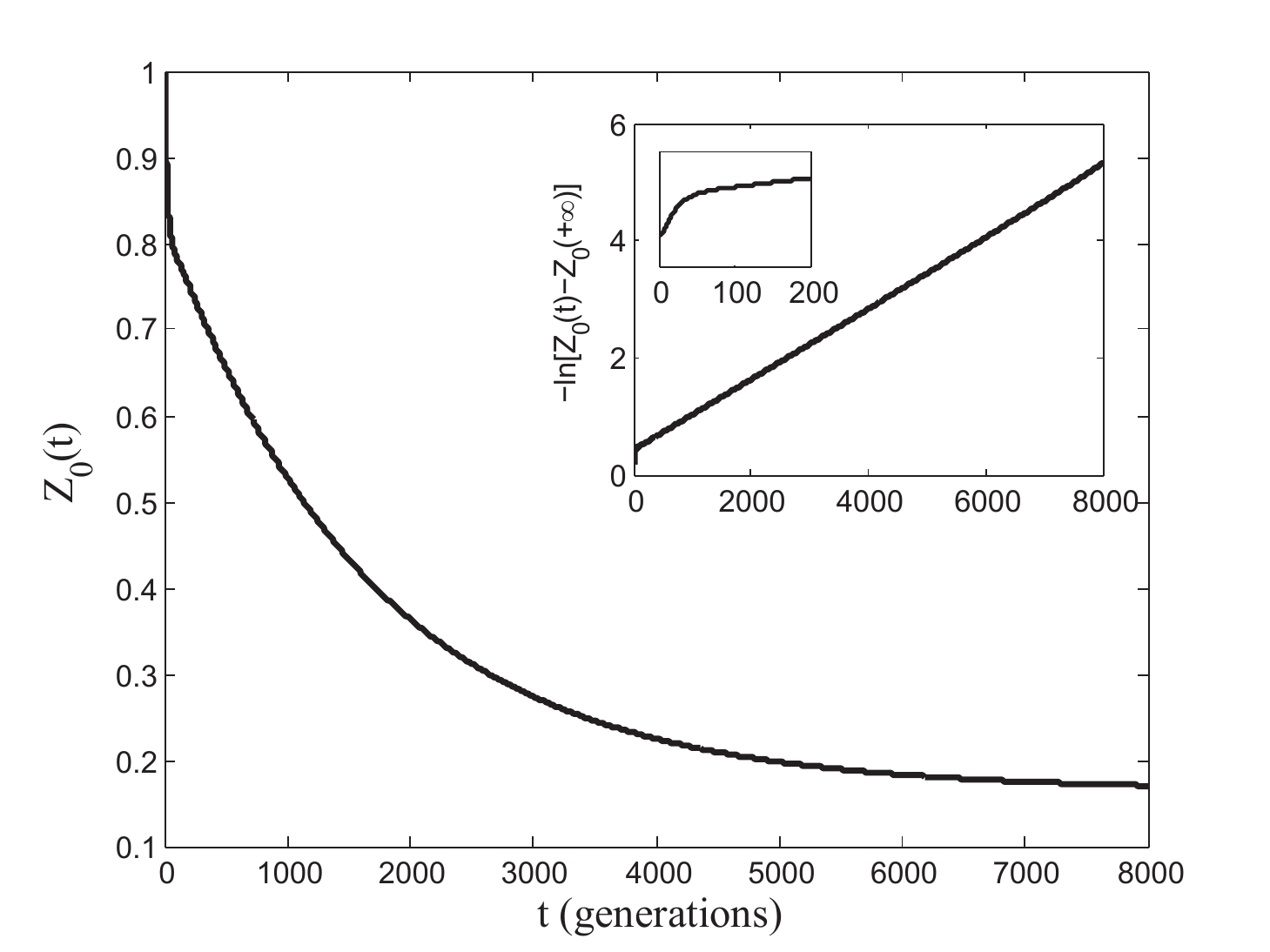}
\caption{Simulation of the escape rate from the attractive basin $(0,a)$ under mutation and drift. Parameter settings: $N=30, \mu=0.0001, \nu=0.0005$. The main plot describes how the cumulative probability density in $(0,a)$ attractive basin $Z_0(t)$ (defined in Eq.~(\ref{Eq:Z0})) changes with time. The inserted figure is the value $-\ln[Z_0(t)-Z_0(+\infty)]$, whose slope gives the flux rate between the two attractive basins. The zoomed-in figure shows the same value in the first timescale. The simulated values are: $\hat{T}_1\approx 62.77 \textrm{(regressing the steady exponential interval $(0,N)$)},~ \hat{T}_2\approx 1464,~\hat{\tau}_0\approx 1755$. Under the same setting, the theoretical expectations are: $T_1= 60,~T_2= 1666,~ \tau_0= 2000$) }
\label{Fig:Z0}
\end{figure}

\begin{figure}[h]
\centering
\includegraphics{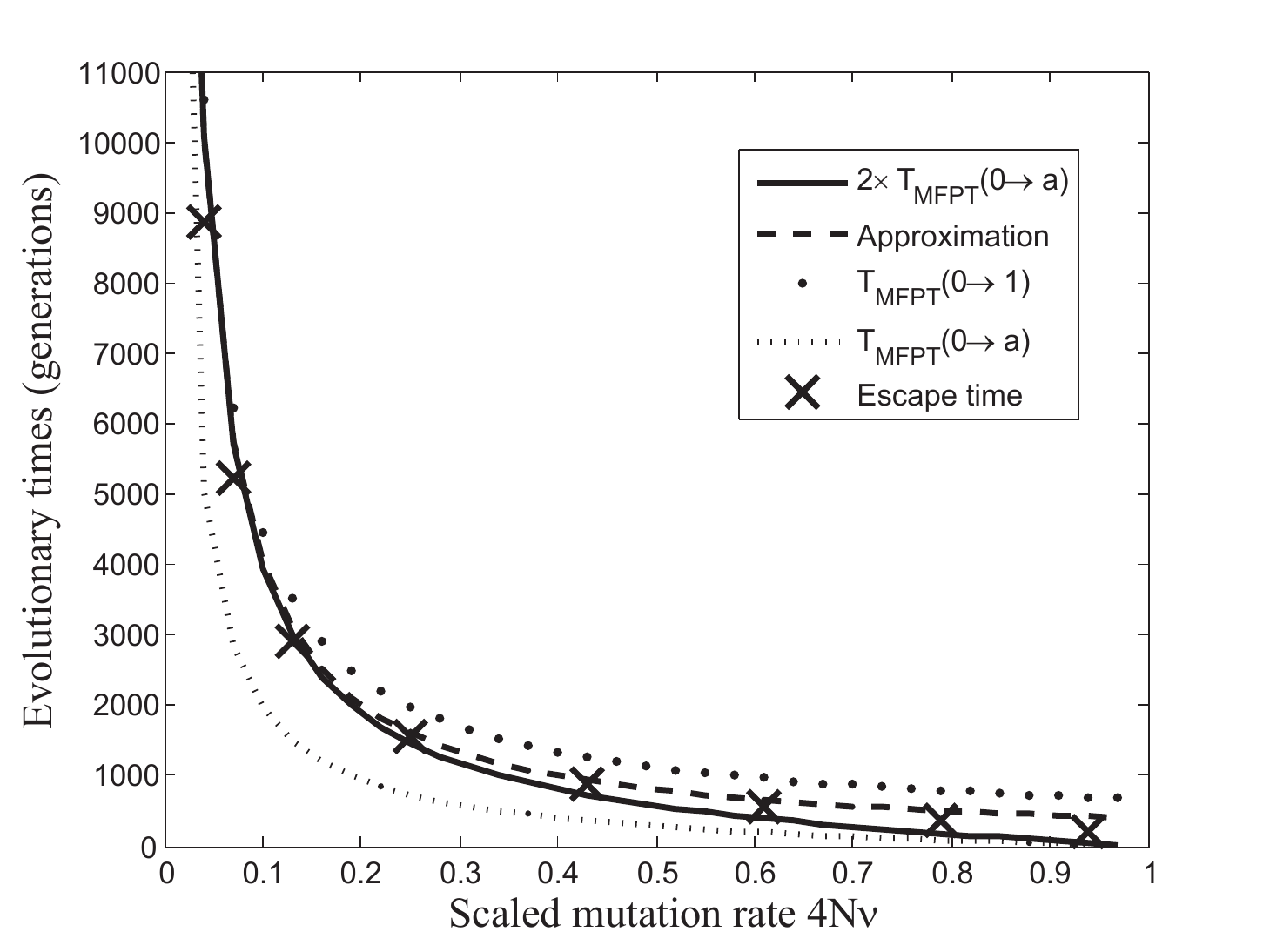}
\caption{Comparison between two times the MFPT from 0 to $a$ (solid), our analytical approximation (dashed), the MFPT from 0 to 1 (point), the MFPT from 0 to $a$ (dotted), and the simulated escape time (crosses) of the discrete Wright-Fisher model under mutation and genetic drift. Parameter settings: $N=100, \mu=0.00005$. }
\label{Fig:Escape_u}
\end{figure}

\begin{figure}[h]
\centering
\includegraphics{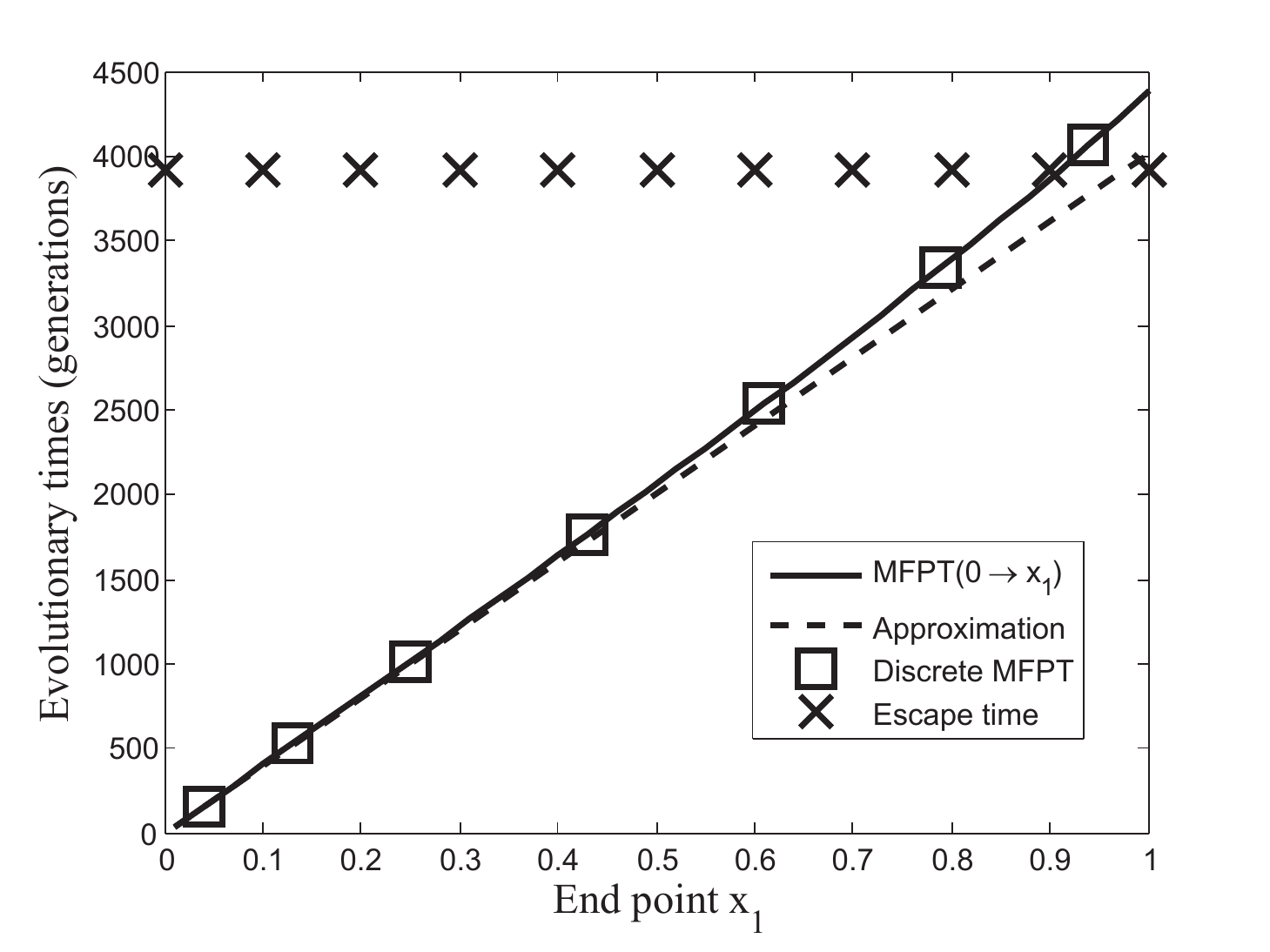}
\caption{Comparisons between the MFPT from 0 to $x_1$ (solid), our analytical approximation of the MFPT (dashed), and the discrete MFPT calculated from the Master equation (square). The parameter setting is $N=100,~\nu=0.00025,~\mu=0.00001$. The saddle point $a=0.4747$. The crosses are the simulated escape time of the discrete Wright-Fisher model under the same setting.}
\label{Fig:MFPT}
\end{figure}

\begin{figure}[h]
\centering
\includegraphics{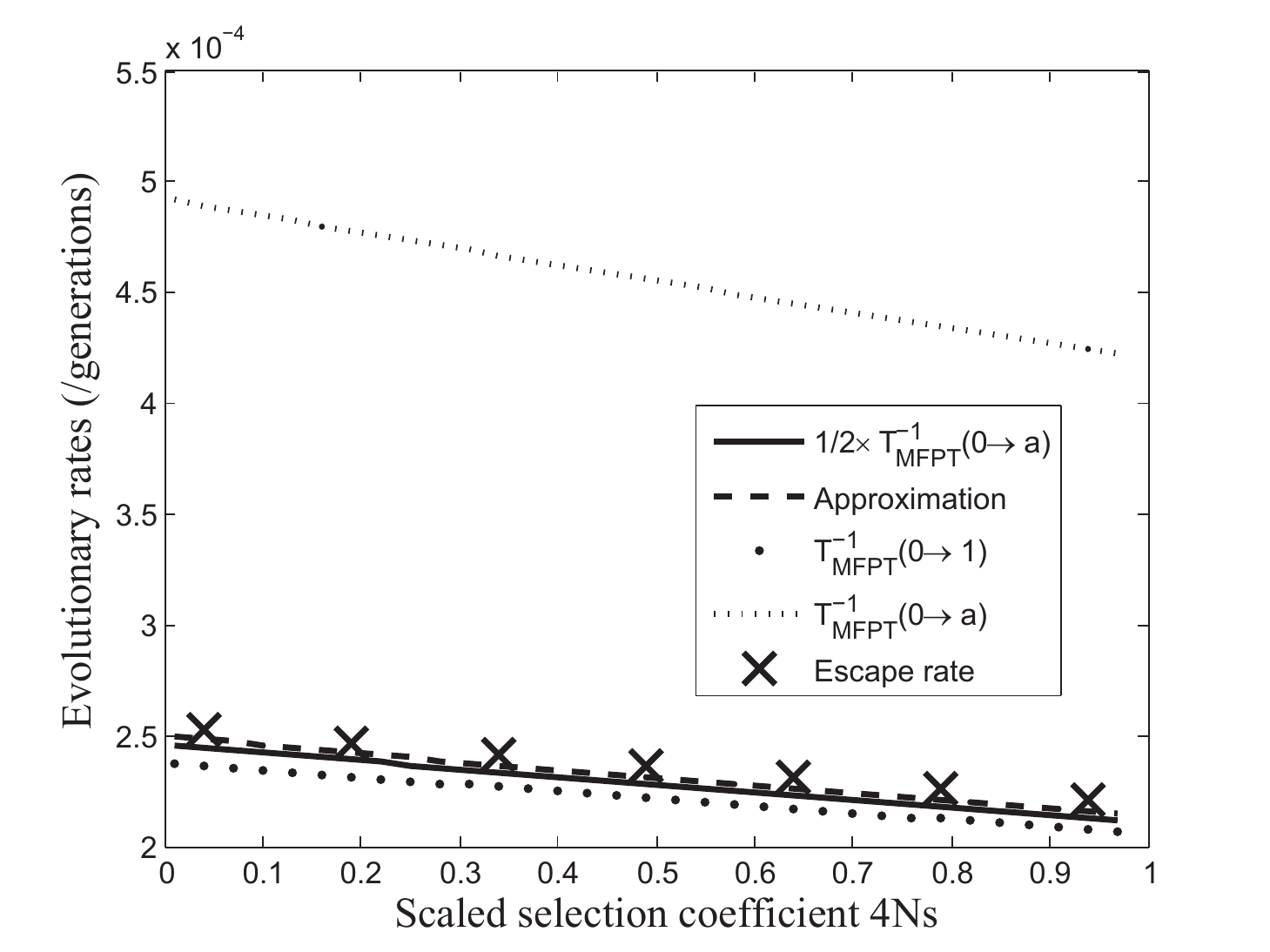}
\caption{Comparison between two times the MFPT from 0 to $a$ (solid), our analytical approximation (dashed), the MFPT from 0 to 1 (point), the MFPT from 0 to $a$ (dotted), and the simulated escape time (crosses) of the discrete Wright-Fisher model under selection, mutation, and genetic drift. Parameter settings: $N=50, \nu=0.00025, \mu=0.00001$. }
\label{Fig:Escape_s}
\end{figure}

\end{document}